\documentclass[%
 reprint,
 superscriptaddress,
 amsmath,amssymb,
 aps,
]{revtex4-1}

\usepackage{graphicx}
\usepackage{dcolumn}
\usepackage{bm}
\usepackage{xcolor}
\usepackage{url}
\usepackage{longtable}

\usepackage{subfigure}
\usepackage{multirow}

\begin{document}


\title{The neutron and proton mass radii from the vector meson
photoproduction data on the deuterium target} 

\author{Chengdong Han}
\affiliation{Institute of Modern Physics, Chinese Academy of Sciences, Lanzhou 730000, China}
\affiliation{University of Chinese Academy of Sciences, Beijing 100049, China}

\author{Gang Xie}
\affiliation{Institute of Modern Physics, Chinese Academy of Sciences, Lanzhou 730000, China}
\affiliation{Guangdong Provincial Key Laboratory of Nuclear Science, Institute of Quantum Matter, South China Normal University, Guangzhou 510006, China}

\author{Wei Kou}
\affiliation{Institute of Modern Physics, Chinese Academy of Sciences, Lanzhou 730000, China}
\affiliation{University of Chinese Academy of Sciences, Beijing 100049, China}

\author{Rong Wang}
\email{rwang@impcas.ac.cn}
\affiliation{Institute of Modern Physics, Chinese Academy of Sciences, Lanzhou 730000, China}
\affiliation{University of Chinese Academy of Sciences, Beijing 100049, China}

\author{Xurong Chen}
\email{xchen@impcas.ac.cn}
\affiliation{Institute of Modern Physics, Chinese Academy of Sciences, Lanzhou 730000, China}
\affiliation{University of Chinese Academy of Sciences, Beijing 100049, China}
\affiliation{Guangdong Provincial Key Laboratory of Nuclear Science, Institute of Quantum Matter, South China Normal University, Guangzhou 510006, China}


\date{\today}

\begin{abstract}
In this study, we try to extract the mass radii of the neutron and the proton
from the differential cross section data of near-threshold $\omega$ and $\phi$
photoproductions on deuterium target, which is often approximated as a quasi-free neutron
plus a quasi-free proton. The incoherent data of $\omega$ and $\phi$ photoproductions
are provided by CBELSA/TAPS collaboration and LEPS collaboration respectively,
where the deuteron is disintegrated in the experiments to measure the properties of individual nucleons.
Under the VMD model and the assumption of dipole gravitational form factor,
we determined the loosely bound neutron and proton mass radii to be
$0.795\pm0.092\rm(stat.)\pm0.073\rm(syst.)$ fm and $0.744\pm0.029\rm(stat.)\pm0.042\rm(syst.)$ fm respectively
from the near-threshold data of $\gamma d \rightarrow \omega n (p)$
and $\gamma d \rightarrow \omega p (n)$, for the first time.
With the near-threshold and incoherent $\phi$ photoproduction data of $\gamma d \rightarrow \phi p n$,
we determined the average mass radius of the bound nucleon (neutron or proton)
inside the deuteron to be $0.755\pm0.039\rm(stat.)\pm0.039\rm(syst.)$ fm, for the first time.
For a comparison study, we also extracted the mass radius of the free proton
from the $\omega$ photoproduction on the hydrogen target by CBELSA/TAPS collaboration.
Based on our analysis results under the assumptions of VMD model
and a low energy QCD theorem, we find that the neutron mass radius
is consistent with the proton mass radius within the current statistical uncertainties,
and that the nuclear modification on the nucleon mass radius is small inside the deuteron.
\end{abstract}

\maketitle


\section{Introduction}
\label{sec:intro}

Understanding the structure of matter has been deep into
the interior of the nucleon: quarks and gluons.
In the modern physical picture, the nucleon is often
described as a ``bag'' full of quarks, antiquarks and gluons,
existing as the most abundant and stable quantum chromodynamics
(QCD) bound state in nature.
The study of the nucleon structure is one important aspect
to understand the strong interaction force.
But as the physicists gradually deepen their understanding
on the inner structure of the nucleon,
more and more puzzles have arisen as well.
For example, the sum of the masses of the valence quarks only accounts
for about one percent of the proton mass.
Most of the mass of the proton comes from the self-interactions of gluons,
but we lack a more specific understanding of this part \cite{Ji:1994av,Ji:1995sv,Ji:2021pys,Ji:2021mtz,Wang:2019mza}.
The mass distribution inside the hadron shows
one important feature of the particular hadron.
Extracting the proton and neutron mass radii is of significance
for the study of the equations of state of the dense nuclear matter,
such as the neutron star \cite{Danielewicz:2002pu,Zhu:2018qpv,Suparti:2017msx}.
In all, understanding the proton mass problem theoretically and experimentally is
a hot topic in the field of high-energy nuclear physics in recent years
\cite{Lorce:2017xzd,Lorce:2021xku,Metz:2020vxd,Liu:2021gco,Yang:2018nqn,He:2021bof}.

From a theoretical point of view, the graviton would be a useful probe
for exploring the mass structure and mechanical properties of the proton.
However, the gravity is dozens of orders of magnitude weaker
than the electromagnetic force, making the interaction between the graviton
and the proton far exceeding the current limit of any experimental techniques \cite{Pagels:1966zza}.
Moreover, due to the color confinement effect of the strong interaction,
the scattering between the bound quark-gluon system and the graviton is
difficult to calculate directly.
Hence we should look for a realistic way to detect the mass structure of the proton.

In astrophysics and cosmology, the study of the distribution of mass in galaxies
gave birth to the hypothesis of the presence of dark matter in the Universe.
Similarly, for the proton, the charged lepton scattering experiments revealed
the spatial distribution of the quarks (``visible'' to photons),
but do not directly probe the spatial distribution of the gluons (``invisible'' to photons).
Gluons play an important role for the proton mass generation \cite{Ji:2021pys,Yang:2018nqn,He:2021bof}.
Therefore, to probe the mass distribution inside the proton,
we need to think of a new probe other than the photon.

\begin{figure}[htp]
\centering
\includegraphics[width=0.4\textwidth]{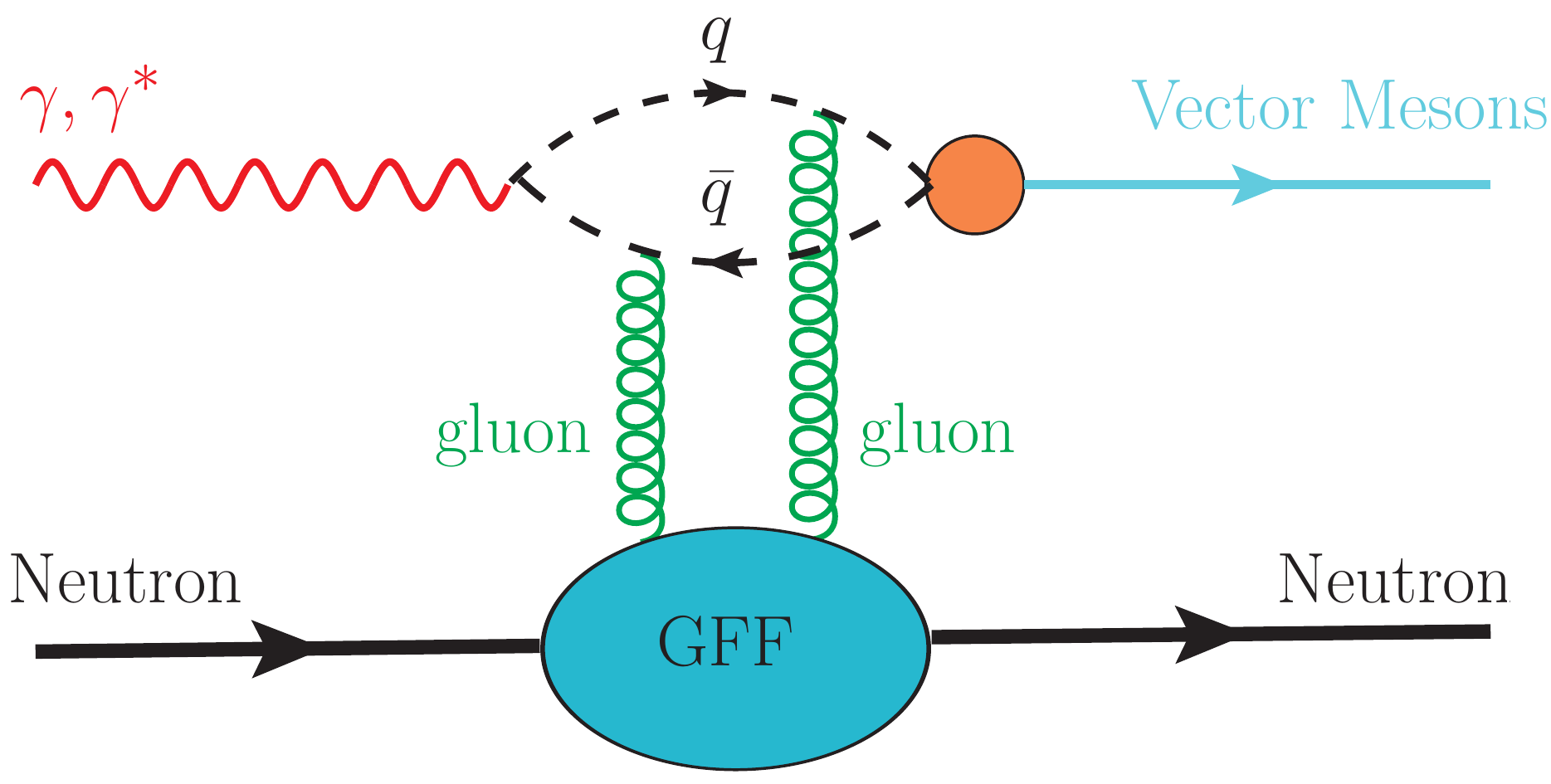}
\caption{
The Feynman diagram of the scattering between the quark-antiquark pair and the neutron
for the near-threshold vector meson photoproductions.
The form factor of the energy-momentum tensor is detected
with the two-gluon exchange process.
}
\label{fig:FeynmanPlot}
\end{figure}

A more feasible way to probe the nucleon mass distribution is to utilize
the elastic scattering between a heavy quarkonium and the hydrogen/deuterium target
\cite{Kharzeev:1995ij,Kharzeev:1998bz,Kharzeev:2021qkd}.
The possible method is to convert the study of graviton-proton scattering
into a scalar gravitational form factor (GFF) of the proton,
under the theoretical framework of vector-meson-dominance (VMD) model \cite{Kharzeev:2021qkd}.
As we focus on the neutron mass radius in this work,
Fig. \ref{fig:FeynmanPlot} shows a diagram of the scattering between a dipole and the neutron.
According to a low-energy theorem, this process of a quarkonium production (vector meson) is
sensitive to the scalar GFF and so to measure the mass distribution inside the neutron \cite{Kharzeev:2021qkd}.

In the weak-field approximation, the scalar GFF can be used
to describe the mass distribution of the nucleon \cite{Kharzeev:2021qkd}.
Specifically, the mass radius $\left<R_{\rm m}^2\right>$ is defined as
the derivative of GFF with zero momentum transfer to the nucleon,
which is given by,
\begin{equation}
\begin{split}
\left<R_{\rm m}^2\right> \equiv \frac{6}{M}\frac{dG(t)}{dt}\big|_{t=0},
\end{split}
\label{eq:MassRadius}
\end{equation}
where the scalar GFF is normalized to $M$ at zero momentum transfer ($G(0)=M$).
This definition is similar to the charge radius,
widely used to estimate the root-mean-square radius
in the process of low-momentum exchange.

The scalar GFF is defined as the form factor of the trace of
QCD energy-momentum tensor (EMT) \cite{Kharzeev:2021qkd,Ji:2021pys,Ji:2021mtz}.
And the scalar GFF is usually parameterized as the dipole form $G(t)=M/(1-t/m_{\rm s}^2)^2$,
with $m_{\rm s}$ a free parameter to be determined by the experimental data.
The differential cross section of quarkonium photoproduction near the threshold is
directly related to the matrix element of the scalar gluon operator,
thus the scalar GFF can be accessed with the process \cite{Kharzeev:2021qkd,Fujii:1999xn}.
In the small momentum transfer region, the differential cross section of
the quarkonium photoproduction near threshold can be described with the scalar GFF,
which is written as the following formalism \cite{Kharzeev:2021qkd,Wang:2021dis,Wang:2021ujy},
\begin{equation}
\begin{split}
\frac{d\sigma}{dt} \propto G^{2}(t).
\end{split}
\label{eq:DiffXsection}
\end{equation}
By analysing the differential cross section of vector meson photoproduction near threshold
on the nucleon target, we could extract the dipole-size parameter $m_{\rm s}$ of the scalar form factor.
Then we can obtain the mass radius information by using Eq. (\ref{eq:MassRadius}).

Tremendous progress has been achieved for the proton structure and radius measurements,
but less is known on the neutron mass radius.
As the proton mass radius has already been fixed by the data
of $\omega$, $\phi$, and J/$\psi$ vector meson photoproductions near threshold
in the previous analyses \cite{Kharzeev:2021qkd,Wang:2021dis,Wang:2021ujy},
we would like to see whether the neutron mass radius
can be extracted in experiment as well.
The recent time-like measurement of neutron shows that
the effective form factor of neutron has the different oscillation behavior
compared to that of proton \cite{BESIII:2021tbq}. It is interesting to see whether or not
the mass distribution and mass radius of the neutron are different
from those of the proton.
To study the neutron mass radius, we could use the deuterium target data,
in which the neutron and the proton are loosely bound.
Therefore we turn to the incoherent photoproduction of vector meson
where the neutron or the proton is knocked out.

The deuterium target is usually taken to study the neutron structure,
as there is no target made of free neutrons in experiment.
However, it is an approximation that the deuteron is viewed as
a free proton and a free neutron.
What we try to extract in this work is actually the mass radius of the quasi-free neutron.
The structure function $F_2$ of nearly free neutron is obtained by the BONuS experiment
with the novel technique of tagging the low-momentum spectator proton in the deuteron \cite{CLAS:2011qvj,CLAS:2014jvt}.
Based on the structure function of nearly free neutron,
the nuclear EMC effect of deuteron is observed \cite{Griffioen:2015hxa},
however it is much weaker than that of the heavy nuclei.
Moreover, the change of nucleon radius due to the nuclear environment
is investigated to be smaller than 3-6\% for the $^3$He nucleus \cite{Sick:1985ygc},
via the analysis of quasi-elastic electron-nucleus scattering.
The modification of nucleon radius inside deuteron should
be even smaller than that of $^3$He.
As nuclear medium effect of deuteron is quite small,
it is reasonable to study the quasi-free neutron mass radius
with the near-threshold $\gamma d \rightarrow V n (p)$ data on the deuterium target,
in which $V$ denotes a vector meson.

\section{Data analyses of vector meson photoproductions near threshold}
\label{sec:data-analysis}

The VMD model is quite successful in describing
the light vector meson photoproductions.
Therefore it provides a valuable approach to study the quarkonium-nucleon
scattering, so as to probe the nucleon mass radius within the theoretical framework
of operator product expansion and low-energy theorems.
Following our previous works assuming a dipole
form GFF to extract the proton and deuteron mass radii from near-threshold vector
meson photoproductions \cite{Wang:2021dis,Wang:2021ujy},
we perform a series of analyses of the neutron and proton mass radii
from the differential cross section data of vector meson photoproductions near thresholds
on the deuterium target, including the experimental data
of $\omega$ and $\phi$ vector mesons \cite{CBELSATAPS:2015wwn,LEPS:2009nuw}.
The CLAS collaboration also reported the incoherent and near-threshold
$\phi$ photoproduction on deuteron target \cite{CLAS:2009kjz,Qian:2010rr}. However,
there are only a few data points reported at $|t|$ larger than 0.5 GeV$^2$,
which are not sensitive to the mass radius of the nucleon.
Therefore, the CLAS data are not included for the extraction
of the quasi-free neutron mass radius.

\subsection{Mass radius of quasi-free neutron with $\omega$ meson probe}
\label{sec:mass-radius-quasifree-neutron}

\begin{figure}[htp]
\centering
\includegraphics[width=0.46\textwidth]{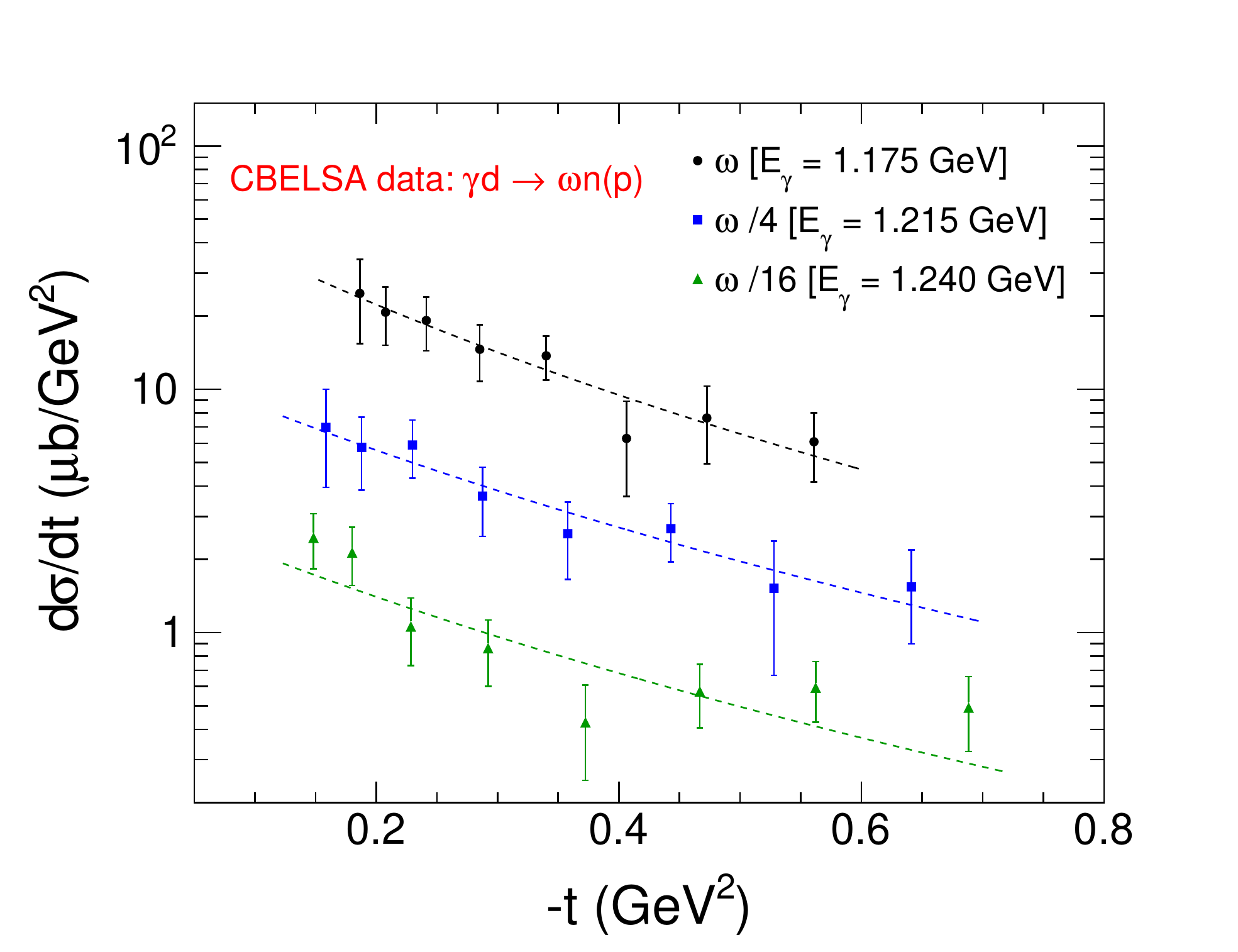}
\caption{
Differential cross sections of the near-threshold photoproduction of $\omega$ meson
as a function of the momentum transfer $-t$ off the quasi-free neutron
in the deuterium target \cite{CBELSATAPS:2015wwn}.
The error bars are statistical only.
The three incident photon energies ($E_{\gamma}=$ 1.175, 1.215 and 1.240 GeV)
near the threshold of $\omega$ meson production are labeled in the figure.
Some cross sections are scaled using the coefficients indicated in the figure,
to avoid the overlapping of the data points.
}
\label{fig:Omega_neutron_Rm}
\end{figure}

\begin{table}[h]
  \caption{
    The extracted values of the dipole-size parameter $m_{s}$ and the quasi-free neutron mass radii $R_{m}^{n^*}$
    from the differential cross-section data of near-threshold $\omega$ productions off the bound neutron
    in deuterium at different incident photon energies. The uncertainties are statistical only.
  }
  \begin{center}
    \begin{ruledtabular}
      \begin{tabular}{ cccc }
        $E_{\gamma}$ (GeV)                       &      1.175         &      1.215          &  1.240        \\
        \hline
        $m_{\rm s}$ (GeV)                        &  $0.799\pm 0.145$  &  $0.893\pm 0.179$   &  $0.900\pm 0.200$    \\
        $\sqrt{\left<R_{\rm m}^2\right>}$ (fm)   &  $0.855\pm 0.155$  &  $0.765\pm 0.153$   &  $0.759\pm 0.169$    \\
      \end{tabular}
    \end{ruledtabular}
  \end{center}
  \label{tab:NeutronRList_Omega}
\end{table}

Fig. \ref{fig:Omega_neutron_Rm} shows the differential cross section of the $\omega$ meson
produced on the bound neutron as a function of the momentum transfer $-t$,
at different photon energies near the threshold.
For the measurement of this reaction at ELSA \cite{CBELSATAPS:2015wwn},
exactly four neutral hits are identified
(three photons from the decay of $\omega$ and one neutron that struck out from deuteron),
to make sure the $\omega$ meson was produced on the bound neutron
in the liquid deuterium target and the exclusivity
($\gamma d \rightarrow \omega n (p)$).
The $t$-dependence of the differential cross section is fitted
with the scalar GFF $G(t)$ of the dipole parametrization.
We determined the dipole parameter $m_{s}$ and the quasi-free neutron mass radius $R_{m}^{n^*}$
from the model fitting to the deuterium data \cite{CBELSATAPS:2015wwn}
at three different incident photon energies ($E_{\gamma}=1.175, 1.215, 1.240$ GeV).
The obtained dipole parameter $m_{s}$ and the quasi-free neutron mass radii $R_{m}^{n^*}$
at different near-threshold energies are listed in Table \ref{tab:NeutronRList_Omega}.
The averaged quasi-free neutron mass radius of the three extracted values
at different energies is calculated to be $0.795 \pm 0.092$(stat.) fm.
We used the following formula for the calculation of weighted average:
$\bar{x}\pm \delta\bar{x}=\sum_i w_i x_i/\sum_i w_i\pm \left(\sum_i w_i\right)^{-1/2}$
with $w_i=1/(\delta x_i)^2$ \cite{ParticleDataGroup:2020ssz}.
Note that the weighted average of the mass radii at different energies
obtained here is consistent with the result of the simultaneous fit
to all the data sets shown in Sec. \ref{sec:ModelDependence}.

\subsection{Mass radius of quasi-free proton with $\omega$ meson probe}
\label{sec:mass-radius-quasifree-proton}

\begin{figure}[htp]
\centering
\includegraphics[width=0.46\textwidth]{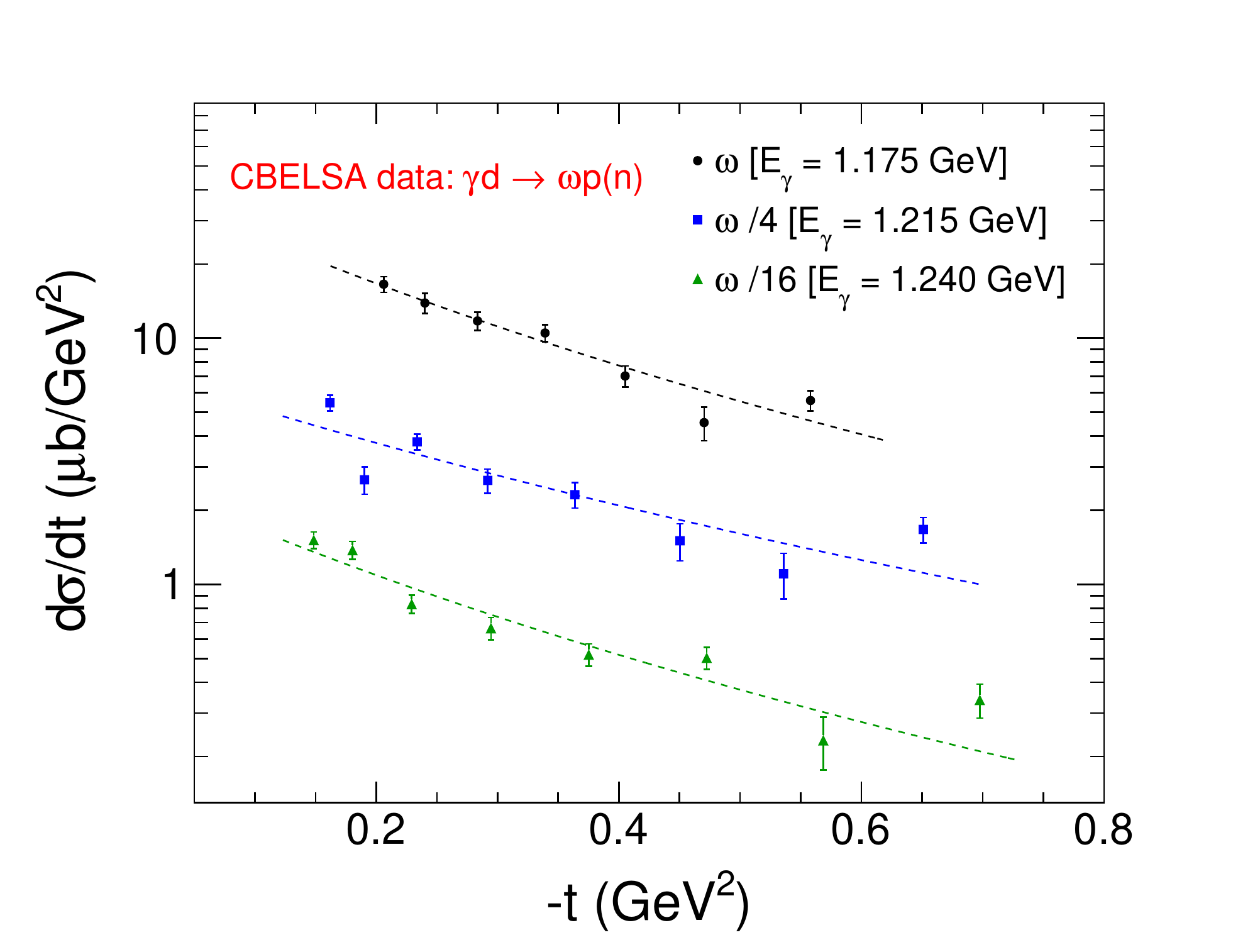}
\caption{
Differential cross sections of the near-threshold photoproduction of $\omega$ meson
as a function of the momentum transfer $-t$ off the quasi-free proton
in the deuterium target \cite{CBELSATAPS:2015wwn}.
The error bars are statistical only.
The three incident photon energies ($E_{\gamma}=$ 1.175, 1.215 and 1.240 GeV)
near the threshold of $\omega$ meson production are labeled in the figure.
Some cross sections are scaled using the coefficients indicated in the figure,
to avoid the overlapping of the data points.
}
\label{fig:Omega_proton_Rm}
\end{figure}

\begin{table}[h]
  \caption{
    The extracted values of the dipole-size parameter $m_{s}$ and the quasi-free proton mass radii $R_{m}^{p^*}$
    from the differential cross-section data of near-threshold $\omega$ productions off the bound proton
    in deuterium at different incident photon energies. The uncertainties are statistical only.
  }
  \begin{center}
    \begin{ruledtabular}
      \begin{tabular}{ cccc }
        $E_{\gamma}$ (GeV)                       &      1.175          &      1.215          &  1.240        \\
        \hline
        $m_{\rm s}$ (GeV)                        &  $0.863\pm 0.056$  &  $1.035\pm 0.076$  &  $0.881\pm 0.055$    \\
        $\sqrt{\left<R_{\rm m}^2\right>}$ (fm)   &  $0.792\pm 0.051$  &  $0.660\pm 0.048$  &  $0.776\pm 0.048$    \\
      \end{tabular}
    \end{ruledtabular}
  \end{center}
  \label{tab:ProtonRList_Omega}
\end{table}

Fig. \ref{fig:Omega_proton_Rm} shows the differential cross section of the $\omega$ meson
produced on the bound proton as a function of the momentum transfer $-t$,
at different photon energies near the threshold.
For the measurement of this reaction at ELSA \cite{CBELSATAPS:2015wwn},
exactly three neutral hits and one charged hit are identified
(three photons from the decay of $\omega$ and one proton that struck out from deuteron),
to make sure the $\omega$ meson was produced on the bound proton
in the liquid deuterium target and the exclusivity
($\gamma d \rightarrow \omega p (n)$).
The $t$-dependence of the differential cross section is fitted
with the scalar GFF $G(t)$ of the dipole parametrization.
We determined the dipole parameter $m_{s}$ and the quasi-free proton mass radius $R_{m}^{p^*}$
from the model fitting to the deuterium data \cite{CBELSATAPS:2015wwn}
at three different incident photon energies ($E_{\gamma}=1.175, 1.215, 1.240$ GeV).
The obtained dipole parameter $m_{s}$ and the quasi-free proton mass radii $R_{m}^{p^*}$
at different near-threshold energies are listed in Table \ref{tab:ProtonRList_Omega}.
The averaged quasi-free proton mass radius of the three extracted values
at different energies is calculated to be $0.741\pm0.028$(stat.) fm.
The weighted average of the mass radii at different energies
obtained here is consistent with the result of the simultaneous fit
to all the data sets shown in Sec. \ref{sec:ModelDependence}.
We see that the mass radii of the loosely bound proton and loosely bound neutron
inside the deuteron agree with each other within the statistical uncertainties.

\subsection{Mass radius of free proton with $\omega$ meson probe}
\label{sec:mass-radius-free-proton}

\begin{figure}[htp]
\centering
\includegraphics[width=0.46\textwidth]{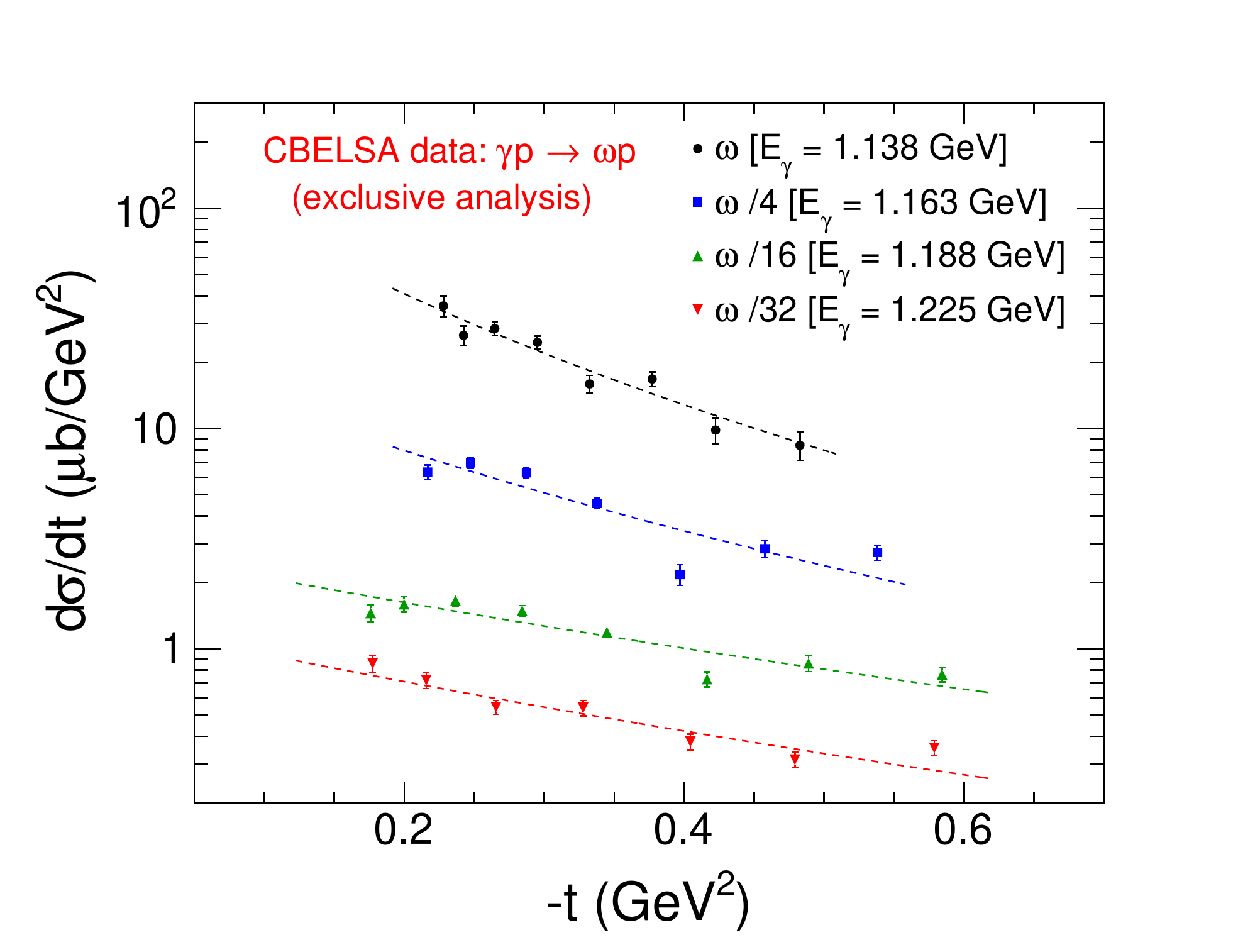}
\caption{
Differential cross sections of the near-threshold photoproduction of $\omega$ meson
as a function of the momentum transfer $-t$ off the free proton
in the hydrogen target \cite{CBELSATAPS:2015wwn}.
The error bars are statistical only.
The four incident photon energies ($E_{\gamma}=$ 1.138, 1.163, 1.188 and 1.225 GeV)
near the threshold of $\omega$ meson production are labeled in the figure.
The exclusive data by CBELSA/TAPS Collaboration are used,
where the recoil proton is identified.
Some cross sections are scaled using the coefficients indicated in the figure,
to avoid the overlapping of the data points.
}
\label{fig:Omega_FreeProton_Rm_Exclusive}
\end{figure}

\begin{table*}[h]
  \caption{
    The extracted values of the dipole-size parameter $m_{s}$ and the free proton mass radii $R_{m}^{p}$
    from the differential cross-section data of near-threshold $\omega$ productions off the free proton
    in hydrogen at different incident photon energies. The data are from the exclusive analysis.
    The uncertainties are statistical only.
  }
  \begin{center}
    \begin{ruledtabular}
      \begin{tabular}{ ccccccccc }
        $E_{\gamma}$ (GeV)                       &      1.138         &      1.163         &  1.188              &  1.225  \\
        \hline
        $m_{\rm s}$ (GeV)                        &  $0.627\pm 0.049$  &  $0.809\pm 0.076$  &  $1.171\pm 0.064$   &  $1.118\pm 0.076$\\
        $\sqrt{\left<R_{\rm m}^2\right>}$ (fm)   &  $1.090\pm 0.084$  &  $0.845\pm 0.079$  &  $0.584\pm 0.032$   &  $0.611\pm 0.042$\\
      \end{tabular}
    \end{ruledtabular}
  \end{center}
  \label{tab:FreeProtonRList_Omega_Exclusive}
\end{table*}

\begin{figure}[htp]
\centering
\includegraphics[width=0.46\textwidth]{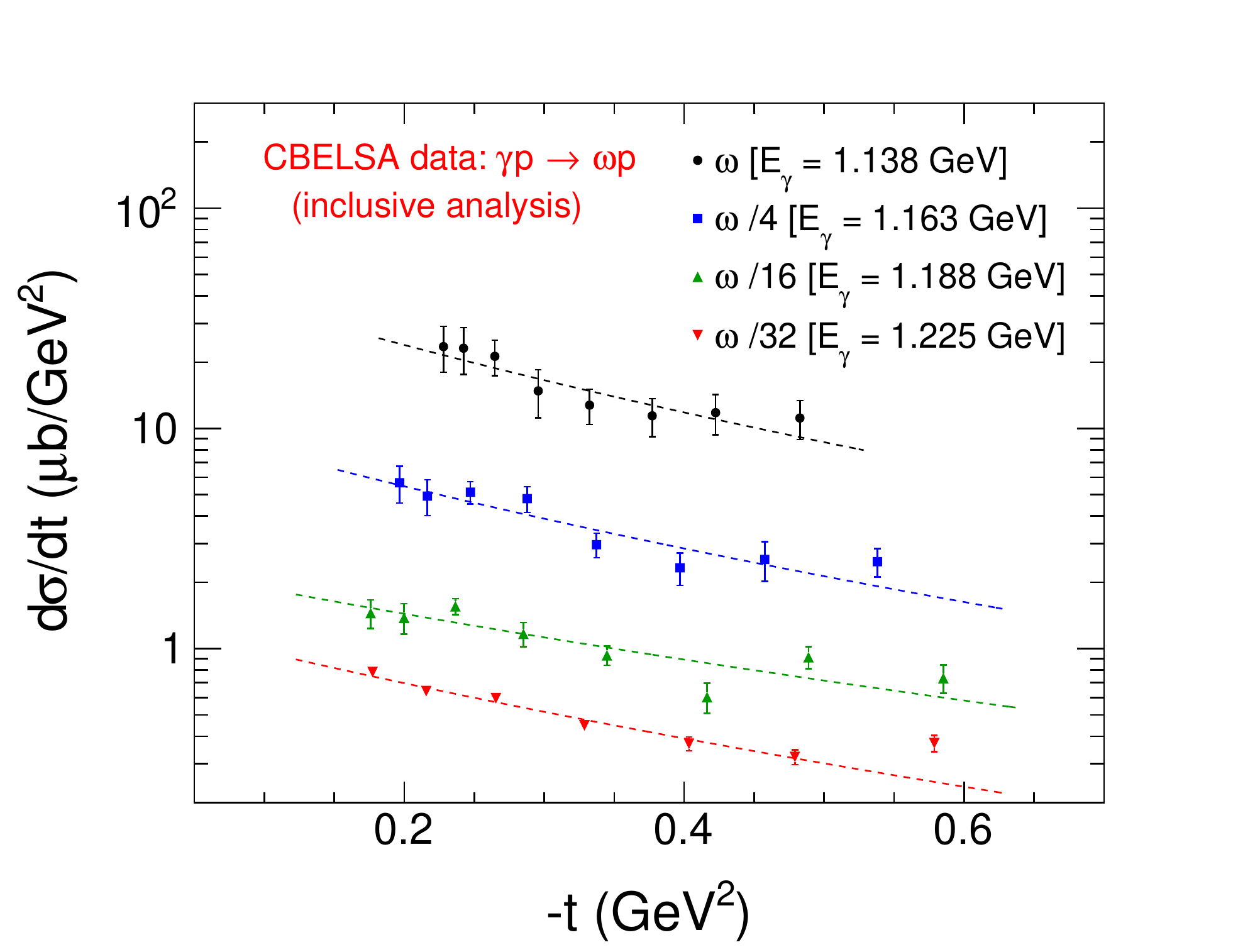}
\caption{
Differential cross sections of the near-threshold photoproduction of $\omega$ meson
as a function of the momentum transfer $-t$ off the free proton
in the hydrogen target \cite{CBELSATAPS:2015wwn}.
The error bars are statistical only.
The four incident photon energies ($E_{\gamma}=$ 1.138, 1.163, 1.188 and 1.225 GeV)
near the threshold of $\omega$ meson production are labeled in the figure.
The inclusive data by CBELSA/TAPS Collaboration are used,
where no requirement is made for the recoil proton detected in coincidence.
Some cross sections are scaled using the coefficients indicated in the figure,
to avoid the overlapping of the data points.
}
\label{fig:Omega_FreeProton_Rm_Inclusive}
\end{figure}

\begin{table*}[h]
  \caption{
    The extracted values of the dipole-size parameter $m_{s}$ and the free proton mass radii $R_{m}^{p}$
    from the differential cross-section data of near-threshold $\omega$ productions off the free proton
    in hydrogen at different incident photon energies. The data are from the inclusive analysis.
    The uncertainties are statistical only.
  }
  \begin{center}
    \begin{ruledtabular}
      \begin{tabular}{ cccccccccc }
        $E_{\gamma}$ (GeV)                       &      1.138         &      1.625         &  1.188              & 1.225\\
        \hline
        $m_{\rm s}$ (GeV)                        &  $0.914\pm 0.207$  &  $0.966\pm 0.128$  &  $1.172\pm 0.127$   &  $1.042\pm 0.044$\\
        $\sqrt{\left<R_{\rm m}^2\right>}$ (fm)   &  $0.748\pm 0.169$  &  $0.708\pm 0.094$  &  $0.583\pm 0.063$   &  $0.656\pm 0.028$\\
      \end{tabular}
    \end{ruledtabular}
  \end{center}
  \label{tab:FreeProtonRList_Omega_Inclusive}
\end{table*}

Fig. \ref{fig:Omega_FreeProton_Rm_Exclusive} and Fig. \ref{fig:Omega_FreeProton_Rm_Inclusive}
show respectively the exclusive and the inclusive cross-section data of
near-threshold $\omega$ meson photoproduction off the free proton in liquid hydrogen target \cite{CBELSATAPS:2015wwn}.
The differential cross sections are shown as a function of momentum transfer $-t$.
For the exclusive data, the recoil nucleon was detected in coincidence.
For the inclusive data, no condition is placed for the detection of the knock-out nucleon.
The $t$-dependence of the differential cross section is fitted
with the scalar GFF $G(t)$ of the dipole parametrization.
We determined the dipole parameter $m_{s}$ and the free proton mass radius $R_{m}^{p}$
from the model fitting to the hydrogen data \cite{CBELSATAPS:2015wwn}
at four different incident photon energies ($E_{\gamma}=1.138, 1.625, 1.188, 1.225$ GeV).
The obtained dipole parameter $m_{s}$ and the free proton mass radii $R_{m}^{p}$
at different near-threshold energies from the exclusive analysis
and the inclusive analysis are listed in Table \ref{tab:FreeProtonRList_Omega_Exclusive}
and Table \ref{tab:FreeProtonRList_Omega_Inclusive}, respectively.
The averaged free proton mass radius of the exclusive
and inclusive analyses are $0.654\pm0.023$(stat.)  fm
and $0.649\pm0.025$(stat.)  fm, respectively,
with the near-threshold $\omega$ production data by CBELSA/TAPS Collaboration \cite{CBELSATAPS:2015wwn}.
The weighted average of the mass radii at different energies
obtained here is consistent with the result of the simultaneous fit
to all the data sets shown in Sec. \ref{sec:ModelDependence}.
We see that the mass radii of the free proton based on
the exclusive data and the inclusive data agree with each other.
We also see that the free proton mass radius may be
smaller than the bound proton mass radius in deuteron,
but more precise data are needed to further check this.

\subsection{The nucleon mass radius with $\phi$ meson probe}
\label{sec:mass-radius-nucleon}

In the following analysis, the nucleon mass radius refers to
the averaged value of the proton mass radius and the neutron mass radius.
From the analyses in above subsections, we have obtained the mass radii
of the quasi-free neutron, the quasi-free proton and the free proton,
with the $\omega$ meson probe.
And the mass radii of the quasi-free neutron and the quasi-free proton are of the similar size.
In this subsection, we would like to check these findings with the measurements of the $\phi$ meson probe.
It is interesting and necessary to look at the probe-dependence
for the neutron mass radius and the proton mass radius.

\begin{figure}[htp]
\centering
\includegraphics[width=0.46\textwidth]{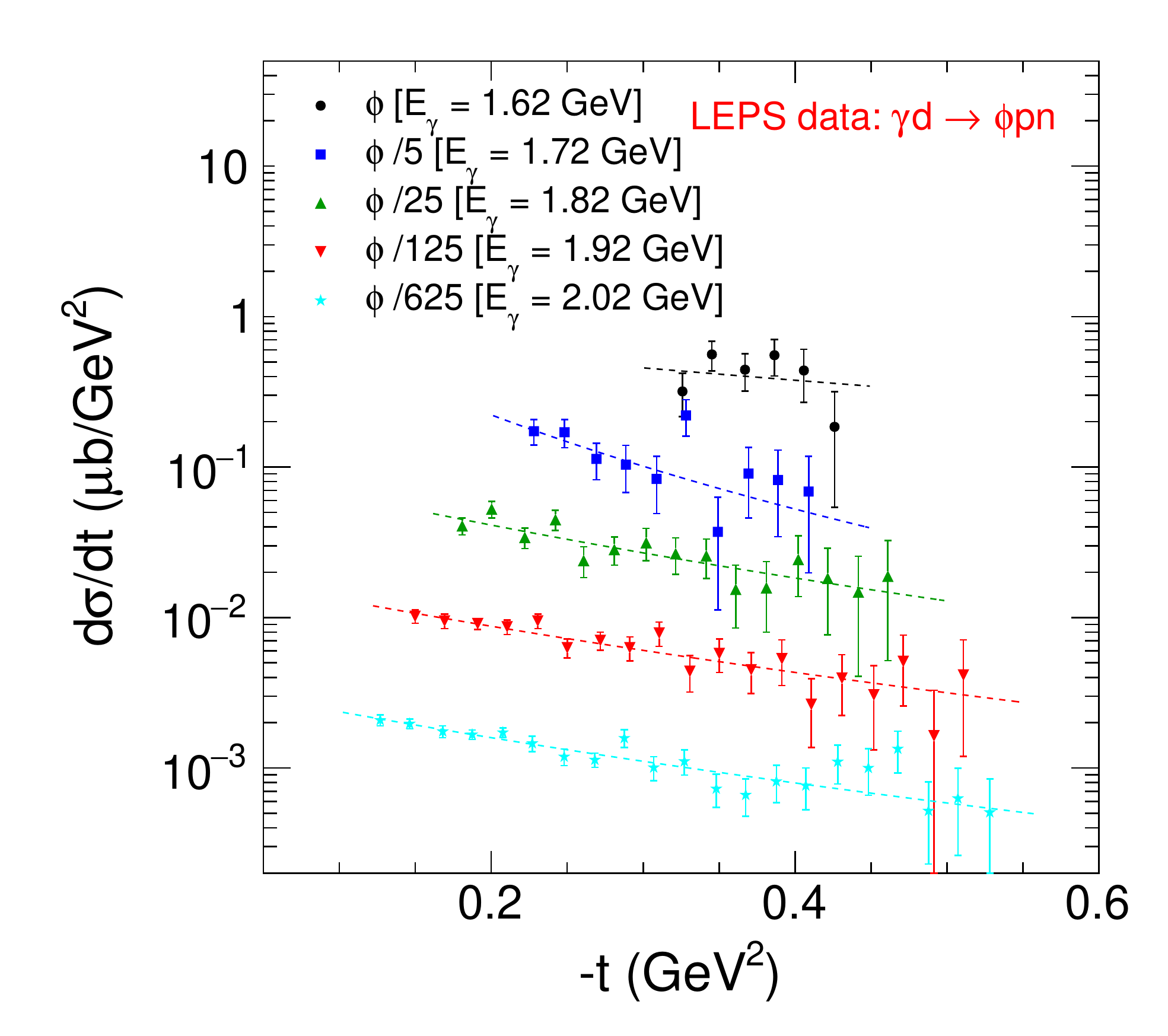}
\caption{
Differential cross sections of the incoherent and near-threshold photoproduction
of $\phi$ meson as a function of the momentum transfer $-t$ off the quasi-free nucleon
in the deuterium target ($\gamma d \rightarrow \phi p n$) \cite{LEPS:2009nuw}.
The error bars are statistical only.
The five incident photon energies ($E_{\gamma}=$ 1.62, 1.72, 1.82, 1.92 and 2.02 GeV)
near the threshold of $\phi$ meson production are labeled in the figure.
Some cross sections are scaled using the coefficients indicated in the figure,
to avoid the overlapping of the data points.
}
\label{fig:Phi_nucleon_Rm}
\end{figure}

\begin{table*}
  \caption{
    The extracted values of the dipole-size parameter $m_{s}$ and the quasi-free nucleon mass radii $R_{m}^{N^*}$
    from the differential cross-section data of incoherent and near-threshold $\phi$ productions
    off the bound nucleon (proton or neutron) in deuterium at different incident photon energies.
    The uncertainties are statistical only.
  }
  \begin{center}
    \begin{ruledtabular}
      \begin{tabular}{ cccccccccc }
        $E_{\gamma}$ (GeV)                       &      1.62         &      1.72         &  1.82              & 1.92                  & 2.02\\
        \hline
        $m_{\rm s}$ (GeV)                        &  $1.326\pm 1.436$  &  $0.511\pm 0.189$  &  $0.831\pm 0.128$   &  $0.914\pm 0.085$  & $0.926\pm 0.064$ \\
        $\sqrt{\left<R_{\rm m}^2\right>}$ (fm)   &  $0.515\pm 0.558$  &  $1.338\pm 0.495$  &  $0.823\pm 0.127$   &  $0.748\pm 0.069$  & $0.738\pm 0.051$ \\
      \end{tabular}
    \end{ruledtabular}
  \end{center}
  \label{tab:RadiusListPhiMeson}
\end{table*}

Fig. \ref{fig:Phi_nucleon_Rm} shows the differential cross section of the $\phi$ meson
produced on the bound nucleon as a function of the momentum transfer $-t$,
at different photon energies near the production threshold \cite{LEPS:2009nuw}.
For the incoherent data on the deuterium target at LEPS \cite{LEPS:2009nuw},
a cut is performed on the missing mass spectra,
to make sure the $\phi$ meson was interacting with the individual nucleon instead of the whole deuteron.
The $t$-dependence of the differential cross section is fitted
with the scalar GFF $G(t)$ of the dipole parametrization.
We determined the dipole parameter $m_{s}$ and the quasi-free nucleon mass radius $R_{m}^{N^*}$
from the model fitting to the deuterium data \cite{LEPS:2009nuw}
at five different incident photon energies ($E_{\gamma}=1.62, 1.72, 1.82, 1.92, 2.02$ GeV).
The obtained dipole parameter $m_{s}$ and the quasi-free nucleon mass radii $R_{m}^{N^*}$
at different near-threshold energies are listed in Table \ref{tab:RadiusListPhiMeson}.
The averaged quasi-free nucleon mass radius of the five extracted values
at different energies is calculated to be $0.752\pm0.039$(stat.) fm, with the $\phi$ meson probe.
The weighted average of the mass radii at different energies
obtained here is consistent with the result of the simultaneous fit
to all the data sets shown in Sec. \ref{sec:ModelDependence}.
This quasi-free nucleon mass radius is just between the mass radii
of the quasi-free neutron and the quasi-free proton probed with the $\omega$ meson probe.
We see that the mass radii of the quasi-free neutron, the quasi-free proton,
and the quasi-free nucleon are consistent with each other.
The dependence on the meson probe is
weak for the mass radius measurement, in terms of the analyses of
the $\omega$ meson data and the $\phi$ meson data.

\section{The model dependence and systematic uncertainties of the analysis}
\label{sec:ModelDependence}

For the determination of the charge radius from electron-proton scattering data,
it is shown that the extracted charge radius depends on the particular functional
form adopted in describing the charge form factor $G_{\rm E}(Q^2)$
and the extrapolation to $Q^2=0$ GeV$^2$ \cite{Barcus:2019skg,Lorenz:2012tm,Hill:2010yb,Yan:2018bez}.
The model dependence of charge radius extraction
on the regression function is shown to be nontrivial, especially
when the very low $Q^2$ data are scarce \cite{Yan:2018bez,Xiong:2019umf}.
For the extraction of mass radius, the model dependence on
the function form in describing the scalar GFF also should be studied.

To investigate the sensitivity of the extracted mass radius
on the function form used in the fit, we performed
the fits of differential cross section data
to three different functional forms:
monopole form $M/(1-t/m_{\rm s}^2)$,
dipole form $M/(1-t/m_{\rm s}^2)^2$
and tripole form $M/(1-t/m_{\rm s}^2)^3$.
In the analysis, we performed the simultaneous fit
to all the experimental data at different energies
with the same slope parameter $m_{\rm s}$
but different normalization parameters.
Tables \ref{tab:BoundNeutronRadii_Omega}, \ref{tab:BoundProtonRadii_Omega},
\ref{tab:FreeProtonRadii_Omega} and \ref{tab:BoundNucleon_Phi}
list the fitting results for
the quasi-free neutron mass radius,
the quasi-free proton mass radius,
the free proton radius and the quasi-nucleon radius, respectively.
We find that the qualities of the fits ($\chi^2/ndf$) are similar
but the extracted radii present some differences.
These results imply that there are some unavoidable uncertainties
from the model assumption for the extrapolation of the data.

\begin{table}[h]
  \caption{
    The mass radii of the quasi-free neutron in deuteron
    extracted from the near-threshold $\omega$ production data on deuterium target ($\gamma d\rightarrow \omega n (p)$),
    based on three different models for scalar gravitational form factor.
    The uncertainties are statistical only.
  }
  \begin{center}
    \begin{ruledtabular}
      \begin{tabular}{ cccc }
        Model                       &      Monopole        &      Dipole          &  Tripole        \\
        \hline
        $\sqrt{\left<R_{\rm m}^2\right>}$ (fm)   &  $1.113\pm0.220$  &  $0.795\pm0.092$   &  $0.733\pm0.075$    \\
                     $\chi^2/ndf$                &       $0.52$      &      $0.61$        &      $0.65$    \\
      \end{tabular}
    \end{ruledtabular}
  \end{center}
  \label{tab:BoundNeutronRadii_Omega}
\end{table}

\begin{table}[h]
  \caption{
    The mass radii of the quasi-free proton in deuteron
    extracted from the near-threshold $\omega$ production data on deuterium target ($\gamma d\rightarrow \omega p (n)$),
    based on three different models for scalar gravitational form factor.
    The uncertainties are statistical only.
  }
  \begin{center}
    \begin{ruledtabular}
      \begin{tabular}{ cccc }
        Model                       &      Monopole        &      Dipole          &  Tripole        \\
        \hline
        $\sqrt{\left<R_{\rm m}^2\right>}$ (fm)   &  $0.965\pm0.058$  &  $0.744\pm0.029$   &  $0.695\pm0.024$    \\
                     $\chi^2/ndf$                &       $3.43$      &      $3.82$        &      $3.97$    \\
      \end{tabular}
    \end{ruledtabular}
  \end{center}
  \label{tab:BoundProtonRadii_Omega}
\end{table}

\begin{table}[h]
  \caption{
    The mass radii of the free proton
    extracted from the near-threshold $\omega$ production data on hydrogen target ($\gamma p\rightarrow \omega p$),
    based on three different models for scalar gravitational form factor.
    The uncertainties are statistical only.
  }
  \begin{center}
    \begin{ruledtabular}
      \begin{tabular}{ cccc }
        Model                       &      Monopole        &      Dipole          &  Tripole        \\
        \hline
        $\sqrt{\left<R_{\rm m}^2\right>}$ (fm)   &  $0.804\pm0.027$  &  $0.667\pm0.016$   &  $0.633\pm0.014$    \\
                     $\chi^2/ndf$                &       $4.70$      &      $4.84$        &      $4.90$    \\
      \end{tabular}
    \end{ruledtabular}
  \end{center}
  \label{tab:FreeProtonRadii_Omega}
\end{table}

\begin{table}[h]
  \caption{
    The mass radii of the quasi-free nucleon in deuteron
    extracted from the near-threshold $\phi$ production data on deuterium target ($\gamma d\rightarrow \phi p n$),
    based on three different models for scalar gravitational form factor.
    The uncertainties are statistical only.
  }
  \begin{center}
    \begin{ruledtabular}
      \begin{tabular}{ cccc }
        Model                       &      Monopole        &      Dipole          &  Tripole        \\
        \hline
        $\sqrt{\left<R_{\rm m}^2\right>}$ (fm)   &  $0.897\pm0.063$  &  $0.755\pm0.039$   &  $0.719\pm0.034$    \\
                     $\chi^2/ndf$                &       $0.86$      &      $0.84$        &      $0.84$    \\
      \end{tabular}
    \end{ruledtabular}
  \end{center}
  \label{tab:BoundNucleon_Phi}
\end{table}

Traditionally the dipole-form parametrization was thought to describe well
the charge form factor of the proton in a wide range of $Q^2$ \cite{Arrington:2007ux,Hofstadter:1956qs},
for the analysis of the old experimental data decades ago.
This is the main reason why the dipole form is employed
in the previous analyses of scalar GFF and proton mass radius \cite{Kharzeev:2021qkd,Wang:2021dis,Wang:2021ujy}.
For the same reason and a direct comparison with the proton charge radius,
the dipole form factor is assumed for the analyses in Sec. \ref{sec:data-analysis}.
In theory, the scalar GFF of dipole form also meets
the recent calculation based on perturbation QCD \cite{Tong:2021ctu}
and the asymptotic behavior predicted by the power counting rule
\cite{Brodsky:1973kr,Matveev:1973ra,Ji:2003fw}.
Therefore the mass radius extracted from dipole form parametrization
is more reliable and suggested.
In experiment, much more and precise data are needed
to differentiate the models used in the fittings.

To reduce the systematic uncertainty from the model assumption
for the extrapolation to $|t|=0$ GeV$^2$,
experimental data of near-threshold vector meson photoproduction
at very small $|t|$ are preferred,
similar to the charge radius measurement via electron-proton elastic scattering.
The low $Q^2$ data reduce the biases from different model assumptions significantly,
for the extracted proton charge radius \cite{Yan:2018bez,Xiong:2019umf}.

Another big part of the systematic uncertainty
comes from the experimental measurement process.
The CBELSA/TAPS Collaboration reported the total systematic uncertainties
to be 15-20\% and 20-35\% for the measurements $\gamma d \rightarrow \omega p (n)$
and $\gamma d \rightarrow \omega n (p)$ respectively \cite{CBELSATAPS:2015wwn}.
These systematic uncertainties in CBELSA/TAPS experiment bring
about the uncertainties of 0.032-0.042 fm and 0.042-0.073 fm
for the extracted quasi-proton and quasi-neutron mass radii respectively,
under the dipole form factor assumption.
The LEPS Collaboration reported that the systematic uncertainties arise from
the disentanglement fit (10-15\%), background (5-10\%), luminosity (5\%) and track
reconstruction efficiency (5-10\%), in the measurement of $\gamma d \rightarrow \phi p n$ \cite{LEPS:2009nuw}.
The total systematic error for the LEPS data are combined to be 14-22\%.
These systematic uncertainties in LEPS experiment bring
about the uncertainty of 0.025-0.039 fm for the extracted nucleon mass radius,
under the dipole form factor assumption.

\begin{figure}[htp]
\centering
\includegraphics[width=0.46\textwidth]{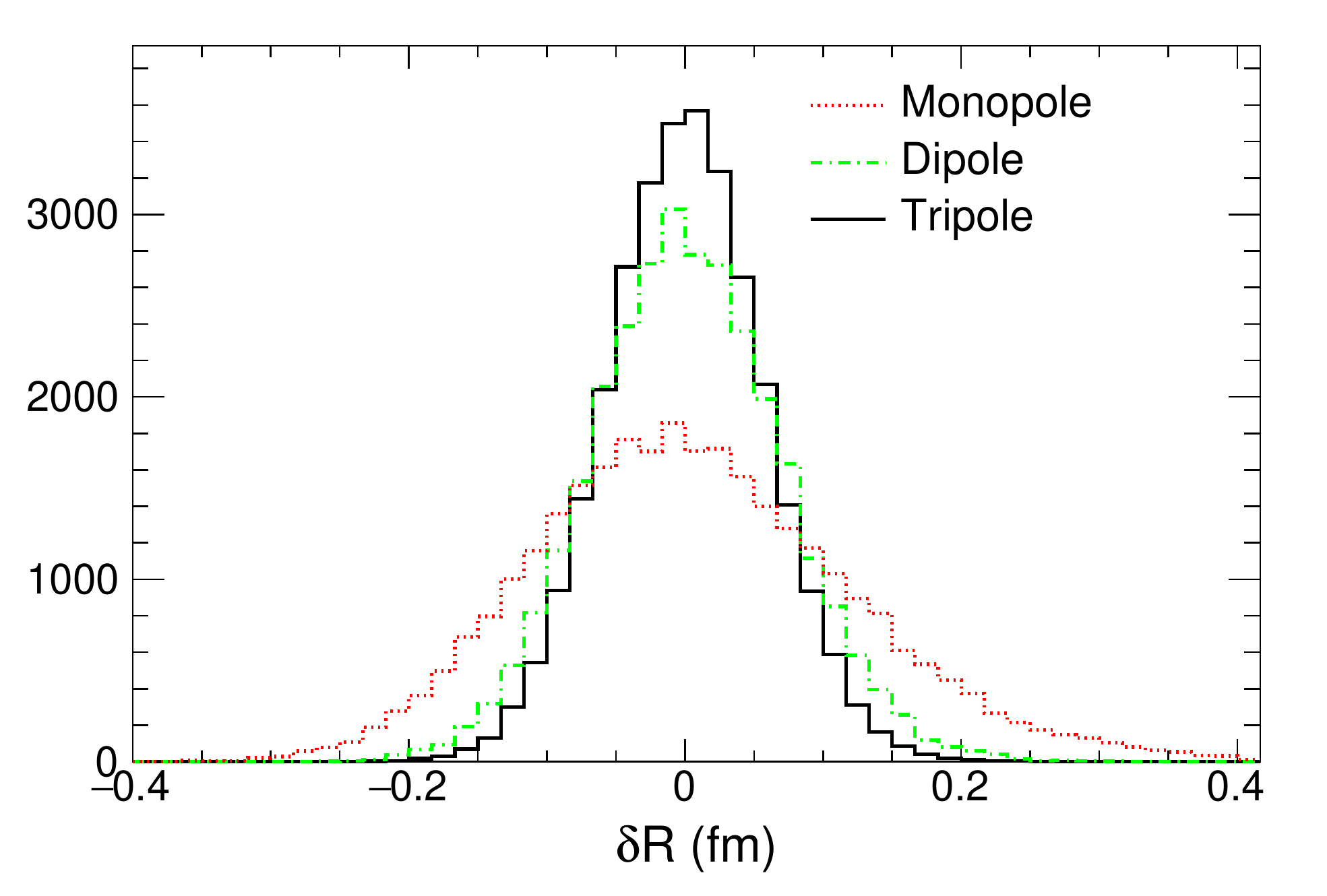}
\caption{The distributions of the bias in fitting process ($\delta R=R^{\rm fit}-R^{\rm input}$)
with different model assumptions for the scalar gravitational form factor,
generated by the MC simulations.
}
\label{fig:model-fitting-bias}
\end{figure}

The minor systematic error would be the intrinsic bias that could be generated
by the fitting model itself. To check this, we did some Monte-Carlo (MC) studies
with monopole-like, dipole-like and tripole-like form factors.
In the MC studies, we generate the gaussian distributed pseudo-data points
according to the statistical errors in experiment.
The input neutron mass radius is 0.75 fm for the MC data.
Some fittings are performed to the pseudo data.
The distributions of the biases (the extracted value minus the input value)
are shown in Fig. \ref{fig:model-fitting-bias}.
The biases by fitting models are estimated to be $0.0046\pm 0.0007$ fm,
$0.0014\pm 0.0004$ fm and $0.0005\pm0.0004$ fm
for monopole-like, dipole-like and tripole-like form factors, respectively.

\section{Discussions and summary}
\label{sec:summary}

\begin{table*}[h]
  \caption{
    The ratio of the quasi-free neutron mass radius $R_{m}^{n^*}$
    to the quasi-free proton mass radius in the deuteron $R_{m}^{p^*}$,
    the ratio of the quasi-free proton mass radius in the deuteron $R_{m}^{p^*}$
    to the free proton mass radius $R_{m}^{p}$,
    and the ratio of the quasi-free neutron mass radius
    in the deuteron $R_{m}^{n^*}$ to the free proton mass radius $R_{m}^{p}$.
  }
  \begin{center}
    \begin{ruledtabular}
      \begin{tabular}{ cccc }
        $Category$         &  $R_{m}^{n^*} / R_{m}^{p^*}$   &  $R_{m}^{p^*} / R_{m}^{p}$     &  $R_{m}^{n^*} / R_{m}^{p}$ \\
        \hline
        $Ratio$            &  $1.069\pm 0.130$                   &  $1.115\pm 0.059$                     &  $1.192\pm 0.062$                    \\
      \end{tabular}
    \end{ruledtabular}
  \end{center}
  \label{tab:RadiusRatioList}
\end{table*}

Based on the extracted mass radii of the bound neutron in deuteron,
the bound proton in deuteron, and the free proton,
we calculated the ratios among them.
In Table \ref{tab:RadiusRatioList}, we list the ratio of the bound neutron mass radius
to the bound proton mass radius, the ratio of the bound proton mass radius in deuteron
to the free proton mass radius, and the ratio of the bound neutron mass radius in deuteron
to the free proton mass radius.
We find that the obtained quasi-free neutron mass radius is about $6.9 \pm 13.0 \%$
larger than the obtained quasi-free proton mass radius.
The obtained bound proton mass radius in deuteron is about $11.5 \pm 5.9 \%$
larger than the free proton mass radius.
To conclude, first, the neutron mass radius is consistent with
the proton mass radius, with the current precision of the experiments.
This is within our expectations, since the isospin symmetry is a rather good symmetry.
Second, the mass radius of the bound nucleon may be a little bit larger
than that of the free nucleon. The result is actually consistent
with the popular ``nucleon swelling'' picture for the explanation
of nuclear medium modifications on parton distribution functions
\cite{Close:1983tn,Jaffe:1982rr,Wang:2018wfz,Miller:2019mae}.

The near-threshold photoproduction of a vector meson on the neutron or the proton
can be well described with the scalar GFF.
Within this model, we determined the neutron mass radius in deuteron to be $0.795\pm0.092$ fm
from the data of $\omega$ photoproductions near the threshold energies.
We also determined the nucleon mass radius in deuteron to be $0.755\pm0.039$ fm
from the incoherent $\phi$ photoproduction data on the deuterium target ($\gamma d \rightarrow \phi p n$).
The obtained neutron and proton mass radii are consistent with the obtained nucleon mass radius.
Similar to the proton, the neutron mass radius is smaller than the neutron magnetic radius.
The possible explanation is that the mass distribution mainly counts on the gluon fluctuations
and the electric current distribution mainly counts on the quarks.

To obtain the mass radius of the nearly free neutron,
the tagging-spectator technique is expected,
like that used in BONuS \cite{CLAS:2011qvj,CLAS:2014jvt}
or ALERT experiment \cite{Armstrong:2017zcm,Armstrong:2017wfw,Armstrong:2017zqr}.
Therefore, we suggest a future experiment of the near-threshold
$\phi$ or J/$\psi$ photoproduction on the deuterium target
with the low-momentum spectator proton tagged,
for the better understanding of the difference between the neutron mass radius
and the proton mass radius.

At present, the Electron-ion collider in China (EicC) \cite{Chen:2018wyz,Chen:2020ijn,Anderle:2021wcy}
and the Electron-Ion Collider in the USA (EIC) \cite{Accardi:2012qut,AbdulKhalek:2021gbh} are proposed,
which will provide good opportunities to study the near-threshold heavy quarkonia photoproductions
on the proton or the nucleus target by exploiting the abundant virtual photon flux at low $Q^2$.
These heavy vector meson photoproduction experiments at EicC and EIC will further test
the VMD model and the scalar GFF assumptions used in this work.
The future experiments at electron-ion colliders will enhance our understanding
on the nucleon mass distribution and radius, which are preliminarily demonstrated and discussed
with the current light quarkonium data.
The study of the mass structure of the nucleon surely will advance our understandings on the
nonperturbative features of QCD, the origins of proton mass,
and the color confinement mechanism of the strong interaction.

\begin{acknowledgments}
This work is supported by the Strategic Priority Research Program of Chinese Academy of Sciences
under the Grant NO. XDB34030301
and the National Natural Science Foundation of China under the Grant NO. 12005266.
\end{acknowledgments}

\bibliographystyle{apsrev4-1}
\bibliography{refs}

\end{document}